\documentclass[12pt]{article}
\usepackage{latexsym}  
\usepackage{amssymb}
\usepackage{graphicx}
\usepackage{amsmath}

\newcommand{\be}{\begin{equation}}
\newcommand{\ee}{\end{equation}}
\newcommand{\bea}{\begin{eqnarray}}
\newcommand{\eea}{\end{eqnarray}}
\newcommand{\bref}[1]{(\ref{#1})}

\newcommand{\bz}{\boldsymbol{\zeta}}
\newcommand{\bomega}{\boldsymbol{\Omega}}
\begin{document}
\begin{titlepage}
\begin{flushright}
\today
\end{flushright}
\vspace{4\baselineskip}
\begin{center}
{\Large\bf Derivation of Generalized Thomas-Bargmann-Michel-Telegdi Equation for a Particle with Electric Dipole Moment}
\end{center}
\vspace{1cm}
\begin{center}
{\large Takeshi Fukuyama$^{a,}$
\footnote{E-mail:fukuyama@se.ritsumei.ac.jp}}
and
{\large Alexander J. Silenko$^{b,c,}$
\footnote{E-mail:alsilenko@mail.ru}}
\end{center}
\vspace{0.2cm}
\begin{center}
${}^{a}$ {\small \it Research Center for Nuclear Physics (RCNP),
Osaka University, Ibaraki, Osaka, 567-0047, Japan}\\[.2cm]

${}^{b} $ {\small \it Institute of Nuclear Problems, Belarusian State University, Minsk 220030, Belarus}

${}^{c} $ {\small \it Bogoliubov Laboratory of Theoretical Physics, Joint Institute for Nuclear Research,
Dubna 141980, Russia}

\vskip 10mm
\end{center}
\vskip 10mm
\begin{abstract}
General classical equation of spin motion is explicitly derived for a particle with magnetic and electric dipole moments in electromagnetic fields. Equation describing the spin motion relative to the momentum direction in storage rings is also obtained.

\medskip


\medskip

\end{abstract}

\end{titlepage}
Electric dipole moments (EDMs) of particles, atoms and molecules are smoking guns of new physics beyond the Standard Model (New Physics BSM).
There are many ongoing and near future experiments measuring EDMs and anomalous magnetic dipole moments (MDMs) of charged particles \cite{fukuyama}. One of the most important equations is the following 
equation of spin motion,
\be
\frac{d\boldsymbol{\zeta}}{dt}=\bomega_s\times \boldsymbol{\zeta},
\ee
where
\be
\bomega_s=-\frac{e}{m}\left[\left(G+\frac{1}{\gamma}\right){\bf B}-\left(G+\frac{1}{\gamma+1}\right){\bf v}\times{\bf E}+\frac{\eta}{2}\left({\bf v}\times {\bf B}+{\bf E}\right)\right].
\label{Nelson1}
\ee
In this equation, $\boldsymbol{\zeta}$ is the rest frame spin pseudovector, $G=(g-2)/2,~g=2\mu m/(es),~\eta=2dm/(es),~\gamma=\epsilon/m$ is the Lorentz factor, $\epsilon$ is the kinetic energy including the rest one,
and $s$ is the spin. We consider the case of $s=1/2$ and use the system of units $\hbar=1,~c=1$.

However, to be curious enough, the explicit derivation of this equation has not been published \cite{Nelson}.
There are several confusions on interpretations of this equation. In the present letter, we derive it explicitly.
It is necessary to take into account that the EDM and MDM terms are not fully symmetrical. The difference is caused by the Dirac (``normal'') magnetic moment, the Thomas precession and the evolution of particle momentum. The last two effects depend on the electric charge of the particle.

In the relativistic theory, the spin pseudovector is not conserved. We should obtain 
an equation of motion for the spin of the moving particle. For that purpose, it is convenient to introduce the spin 4-pseudovector and the momentum 4-vector, $a^\mu$ and $p^\mu$, whose definition in the particle rest frame is
given by \cite{Landau}
\be
a^\mu=(0, \boldsymbol{\zeta}), ~~p^\mu=(m,{\bf 0}).
\ee
So, in any frame
\be
a^\mu p_\mu=0, ~~a_\mu a^\mu=-\boldsymbol{\zeta}^2.
\ee
In a frame moving with velocity ${\bf v}={\bf p}/\epsilon$, the $a^\mu=(a^0,~{\bf a})$ 4-pseudovector is defined by
\be
a^\mu=(a^0,~{\bf a}),~~{\bf a}=\boldsymbol{\zeta}+\frac{{\bf p}(\boldsymbol{\zeta}\cdot {\bf p})}{m(\epsilon+m)},~~a^0=\frac{{\bf a}\cdot{\bf p}}{\epsilon}=\frac{{\bf p}\cdot\boldsymbol{\zeta}}{m},~~{\bf a}^2=\boldsymbol{\zeta}^2+\frac{({\bf p}\cdot \boldsymbol{\zeta})^2}{m^2}.
\label{spin}
\ee
The relativistic equation of spin motion in an electromagnetic field using this 4-pseudovector has the form
\be
\frac{d a^\mu}{d\tau}=\alpha F^{\mu\nu}a_\nu+\beta u^\mu F^{\nu\lambda}u_\nu a_\lambda+\gamma F^{*\mu\nu}a_\nu+\delta u^\mu F^{*\nu\lambda}u_\nu a_\lambda
\label{BMT}
\ee
with $F^{*\mu\nu}=\frac{1}{2}\epsilon^{\mu\nu\rho\sigma}F_{\rho\sigma}$.
Here $\alpha,~\beta,~\gamma,~\delta$ are coefficients whose meanings are determined as follows.
In the rest frame, Eq. \bref{BMT} becomes
\be
\frac{d a^i}{dt}=\frac{d\zeta^i}{dt}=\alpha F^{ij}\zeta_j+\gamma F^{*ij}\zeta_j=\alpha (\boldsymbol{\zeta}\times {\bf B})^i+\gamma ({\bf E}\times \boldsymbol{\zeta})^i.
\label{NRspin}
\ee
In this frame, the equation of spin motion is
\be
\frac{d\boldsymbol{\zeta}}{dt}=2\mu {\boldsymbol{\zeta}}\times {\bf B}+2d\boldsymbol{\zeta}\times {\bf E}.
\ee
Comparing this equation with Eq. \bref{NRspin}, we obtain
\be
\alpha=2\mu, ~~\gamma=-2d.
\ee
The value of $\beta$ results from the equation of motion
\be
m\frac{d u^{\mu}}{d\tau}=eF^{\mu\nu}u_\nu
\label{Lorentz}
\ee
and from $a_\mu u^\mu=0$ that
\be
u_\mu\frac{da^\mu}{d\tau}=-a_\mu\frac{du^\mu}{d\tau}=\frac{e}{m}F^{\mu\nu}u_\mu a_\nu.
\ee
On the other hand, multiplying Eq. \bref{BMT} by $u_\mu$ and taking $u_\mu u^\mu=1$ into account, we obtain
\be
u_\mu\frac{da^\mu}{d\tau}=(2\mu+\beta) F^{\mu\nu}u_\mu a_\nu+(-2d+\delta)F^{*\mu\nu}u_\mu a_\nu.
\ee
Then
\be
\beta=-2\left(\mu-\frac{e}{2m}\right)\equiv -2\mu ',~~\delta=2d.
\ee

As a result, the equation of spin motion takes the form
\be
\frac{da^\mu}{d\tau}=2\mu F^{\mu\nu}a_\nu-2\mu'u^\mu F^{\nu\lambda}u_{\nu}a_\lambda-2d(F^{*\mu\nu}a_\nu-u^\mu F^{*\nu\lambda}u_\nu a_\lambda).
\label{BMT1}
\ee
This is the Thomas-Bargmann-Michel-Telegdi (T-BMT) equation \cite{Thomas,Thomas1,BMT} added by the EDM terms.

In the original paper of Bargmann, Michel, and Telegdi \cite{BMT}, an extension of the equation of spin motion due to an electric dipole moment
has already been discussed. However, such an extension has been based on a dual transformation of $F^{\mu\nu}$ into $F^{*\mu\nu}$ and an explicit derivation of Eq. (\ref{BMT1}) has not been presented.

The spatial part of this equation is presented by 
\bea
\frac{d{\bf a}}{dt}&=&\left[\frac{d{\bf a}}{dt}\right]_{MDM}+\left[\frac{d{\bf a}}{dt}\right]_{EDM},\nonumber\\
\left[\frac{d{\bf a}}{dt}\right]_{MDM}&=&\frac{2}{\gamma}\left[\mu\left\{{\bf a}\times {\bf B}+({\bf a}\cdot{\bf v}){\bf E}\right\}+\mu'\gamma^2{\bf v}\left\{-{\bf a}\cdot {\bf E}+{\bf v}\cdot({\bf a}\times{\bf B})+({\bf a}\cdot{\bf v})({\bf v}\cdot{\bf E})\right\}\right],\nonumber\\
\left[\frac{d{\bf a}}{dt}\right]_{EDM}&=&-\frac{2d}{\gamma}\left[({\bf a}\cdot {\bf v}){\bf B}-{\bf a}\times {\bf E}+\gamma^2{\bf v}\left\{-{\bf a}\cdot {\bf B}-{\bf v}\cdot ({\bf a}\times {\bf E})+({\bf a}\cdot{\bf v})({\bf v}\cdot{\bf B})\right\}\right],
\label{BMT2}
\eea
where $\gamma=\epsilon/m$.
We consider the evolution of $\bz$. Since
$$\boldsymbol{\zeta}={\bf a}-\frac{{\bf v}({\bf a}\cdot {\bf v})\epsilon}{\epsilon+m},$$
it is defined by
\bea
\frac{d\boldsymbol{\zeta}}{dt}&=&\left[\frac{d\boldsymbol{\zeta}}{dt}\right]_{MDM}+\left[\frac{d\boldsymbol{\zeta}}{dt}\right]_{EDM},\nonumber\\
\left[\frac{d\boldsymbol{\zeta}}{dt}\right]_{MDM}&=&\left[\frac{d{\bf a}}{dt}\right]_{MDM}-\frac{{\bf v}\epsilon}{\epsilon+m}\left(\left[\frac{d{\bf a}}{dt}\right]_{MDM}\cdot {\bf v}\right)-
\frac{({\bf a}\cdot {\bf v})\epsilon}{\epsilon+m}\cdot\frac{d{\bf v}}{dt}\nonumber\\
&-&\frac{{\bf v}\epsilon}{\epsilon+m}\left({\bf a}\cdot\frac{d{\bf v}}{dt} \right)-\frac{{\bf v}({\bf a}\cdot {\bf v})m}{(\epsilon+m)^2}\cdot\frac{d\epsilon}{dt},\nonumber\\
\left[\frac{d\boldsymbol{\zeta}}{dt}\right]_{EDM}&=&\left[\frac{d{\bf a}}{dt}\right]_{EDM}-\frac{{\bf v}\epsilon}{\epsilon+m}\left(\left[\frac{d{\bf a}}{dt}\right]_{EDM}\cdot {\bf v}\right).
\label{timedev}
\eea

Evidently, the quantity $\left[\frac{d\boldsymbol{\zeta}}{dt}\right]_{MDM}$ is expressed by the T-BMT equation. When we use the decomposition of the equation of motion \bref{Lorentz} into spatial and temporal components,
\be
\frac{d{\bf v}}{dt}=\frac{e}{m\gamma}\left[{\bf E}+{\bf v}\times {\bf B}-{\bf v}({\bf v}\cdot{\bf E})\right],~~\frac{d\epsilon}{dt}=e{\bf v}\cdot {\bf E},
\label{eqm}
\ee
tedious but simple calculations result in
\bea
\left[\frac{d\boldsymbol{\zeta}}{dt}\right]_{MDM}&=&\frac{2\mu m+2\mu'(\epsilon-m)}{\epsilon}\boldsymbol{\zeta}\times {\bf B}+\frac{2\mu'\epsilon}{\epsilon +m}({\bf v}\cdot{\bf B})({\bf v}\times \boldsymbol{\zeta})+\frac{2\mu m+2\mu'\epsilon}{\epsilon+m}\boldsymbol{\zeta}\times ({\bf E}\times {\bf v})\nonumber\\
&=&\frac{e}{m}\left[\left(G+\frac{1}{\gamma}\right)\boldsymbol{\zeta}\times {\bf B}+\frac{G\gamma}{\gamma+1}({\bf v}\cdot{\bf B}){\bf v}\times \boldsymbol{\zeta}+\left(G-\frac{1}{\gamma+1}\right)\boldsymbol{\zeta}\times({\bf E}\times {\bf v})\right].
\label{bmt3}
\eea

We present a more detailed derivation of contribution of the EDM. Substituting the relation ${\bf a}=\boldsymbol{\zeta}+{\bf v}(\boldsymbol{\zeta}\cdot {\bf v})\gamma^2/(\gamma+1)$
into Eq. \bref{BMT2} results in
\bea
&&\left[\frac{d{\bf a}}{dt}\right]_{EDM}=2d\left[\frac{1}{\gamma}\left(\bz\times
{\bf E}\right)+\frac{\gamma}{\gamma+1}({\bf v}\times {\bf E})(\bz\cdot{\bf v})-(\bz\cdot{\bf v}){\bf B}+
\gamma{\bf v}(\bz\cdot{\bf B})\right.\nonumber\\
&&\left. -\frac{\gamma^2}{\gamma+1}{\bf v}({\bf v}\cdot{\bf B})(\bz\cdot{\bf v})+\gamma{\bf v}({\bf v}\cdot
(\bz\times{\bf E}))\right].
\eea

Since
\be
-\frac{{\bf v}\epsilon}{\epsilon+m}\left(\left[\frac{d{\bf a}}{dt}\right]_{EDM}\cdot {\bf v}\right)=-2d\frac{\gamma{\bf v}}{\gamma+1}\left[\gamma({\bf v}\cdot(\bz\times{\bf E}))
+\frac{\gamma^2-1}{\gamma}(\bz\cdot{\bf B})-\gamma(\bz\cdot{\bf v})({\bf v}\cdot{\bf B})\right],
\label{HFW}
\ee
the result is given by
\bea
\left[\frac{d\boldsymbol{\zeta}}{dt}\right]_{EDM}&=&2d\left[\boldsymbol{\zeta}\times {\bf E}+\frac{\gamma}{\gamma+1}({\bf v}\cdot{\bf E})({\bf v}\times \boldsymbol{\zeta})-\boldsymbol{\zeta}\times ({\bf B}\times {\bf v})\right]\nonumber\\
&=&\frac{e\eta}{2m}\left[\boldsymbol{\zeta}\times {\bf E}+\frac{\gamma}{\gamma+1}({\bf v}\cdot{\bf E})({\bf v}\times \boldsymbol{\zeta})-\boldsymbol{\zeta}\times ({\bf B}\times {\bf v})\right].
\label{bmt4}
\eea

The resulting angular velocity of spin precession is
\bea
\bomega_s &=&-\frac{e}{m}\left[\left(G+\frac{1}{\gamma}\right){\bf B}-\frac{\gamma G}{\gamma+1}({\bf v}\cdot{\bf B}){\bf v}-\left(G+\frac{1}{\gamma+1}\right){\bf v}\times{\bf E}\right.\nonumber\\
&+&\left.\frac{\eta}{2}\left({\bf E}-\frac{\gamma}{\gamma+1}({\bf v}\cdot{\bf E}){\bf v}+{\bf v}\times {\bf B}\right)\right].
\label{Nelsonh}
\eea

Equation (\ref{Nelsonh}) coincides with that obtained by Nelson et al. \cite{Nelson} with the use of dual transformation of terms proportional to $G$. The same dual transformation
$G\rightarrow\eta,~{\bf B}\rightarrow{\bf E},~{\bf E}\rightarrow-{\bf B}$ has been applied by Khriplovich \cite{Khriplovich} for the derivation of this equation.
However, the problem is not too simple.
The three-component spin $\bz$ is defined in the \emph{rest frame},
whereas $\bf E$ and $\bf B$ are the fields in the \emph{laboratory frame}. In addition, one also needs to take into account the nontrivial effect of the Thomas precession \cite{Thomas,Thomas1}. We suppose that the explicit derivation of Eq. (\ref{Nelsonh}) which is basic for all EDM experiments is necessary.

This equation also coincides with classical limits of quantum-mechanical equations obtained for spin-1/2 and spin-1 particles in Refs. \cite{NPR,RPJ} and \cite{PRDspin1}, respectively.

The experimental situation usual for EDM experiments in storage rings consists in
\be
{\bf B}\cdot{\bf v}=0,~~{\bf E}\cdot{\bf v}=0.
\label{transversal}
\ee
In this case, Eq. (\ref{Nelsonh}) takes the form (\ref{Nelson1}).

In relation to the EDM experiments in storage rings, taking into account a longitudinal magnetic field 
may be important for calculations of systematical errors.
A longitudinal electric field is usually associated with a beam acceleration. This field can be applied in the EDM experiments performed by the resonance method \cite{OMS}.

One usually considers the spin motion relative to the beam direction and rewrites Eq. (\ref{eqm}) in terms of the unit vector in direction of the velocity (momentum), ${\bf N}={\bf v}/v={\bf p}/p$:
$$\frac{d{\bf
N}}{dt}=\frac{\dot{\bf v}}{v}
-\frac{\bf v}{v^3}\left({\bf v}\cdot\dot{\bf v}\right)=\bomega_p\times{\bf N}, ~~~ \bomega_p=\frac{e}{m\gamma}\left(\frac{{\bf
N}\times{\bf E}}{v}-{\bf B}\right),
$$ where $\bomega_p$ is the angular
velocity of rotation of the velocity, momentum, and beam directions. Thus, the angular velocity of the spin rotation relative to the beam direction is given by
\be
\bomega=\bomega_s-\bomega_p=-\frac{e}{m}\left[G{\bf B}-\left(G-\frac{1}{\gamma^2-1}\right){\bf v}\times{\bf E}+\frac{\eta}{2}\left({\bf E}+{\bf v}\times {\bf B}\right)\right].
\label{Nelsonf}
\ee

For more detailed derivations, especially for calculations of systematical errors, it is helpful to use the cylindrical coordinates \cite{PRStAcBeam}.

Thus, the general classical equation of spin motion has been explicitly derived for the particle with magnetic and electric dipole moments in electromagnetic fields. This equation coincides with the classical limit of
corresponding quantum-mechanical equations previously obtained for spin-1/2 and spin-1 particles in Refs. \cite{NPR,RPJ} and \cite{PRDspin1}, respectively. We also present the equation of spin motion relative to the momentum direction in storage rings.

\section*{Acknowledgements}

We are grateful to J. Pretz and E.D. Commins for valuable arguments.
The work of T.F. is supported in part by the Grant-in-Aid for Science
Research from the Ministry of Education, Science and Culture of Japan
(No. 21104004).
The work of A.S. is supported by the
Belarusian Republican Foundation for Fundamental Research
(Grant No. $\Phi$12D-002).

\end{document}